\documentclass[authoryear,preprint,3p]{elsarticle}
\usepackage[utf8]{inputenc}

\usepackage[authoryear]{natbib}
\usepackage{amssymb, bm}
\usepackage{amssymb,amsthm,epsfig}
\usepackage[centertags]{amsmath}

\newcommand{\vvec}{\textrm{vec}}

\begin{document}

\begin{frontmatter}
\title{A note on Influence diagnostics in nonlinear mixed-effects elliptical models}
\author[A]{Alexandre G.~Patriota\corref{cor1}}
\ead{patriota.alexandre@gmail.com and patriota@ime.usp.br}
\address[A]{Institute of Mathematics and Statistics, University of S\~ao Paulo, S\~ao Paulo/SP, 05508-090, Brazil}
\cortext[cor1]{Corresponding author}
\begin{abstract}
This paper provides general matrix formulas for computing the score function, the (expected and observed)
Fisher information and the $\bm{\Delta}$ matrices (required for the assessment of local 
influence) for a quite general model which includes the one proposed by \cite{Russoetal2009}. Additionally, we also present an expression for the generalized leverage. 
 The matrix formulation has a considerable advantage, since although the complexity of the postulated model, 
all general formulas are compact, clear and have nice forms.
\end{abstract}


\end{frontmatter}

\section{Main results}

Recently, \cite{Russoetal2009} introduce an interesting nonlinear mixed model considering
an elliptical distribution for the response variable. The authors also present 
a motivating example in a kinetics longitudinal data set which was firstly presented 
in \cite{Vonesh1992} and previously analyzed under the assumption of normality. 
\cite{Russoetal2009} analyze this dataset considering heavy-tailed distributions which may accommodate
 ``large'' observations. The authors compute the score function, Fisher information and 
some influence measures, but some matrices are presented  only with the ($r,s$) element. 
 The first author to compute expressions for the entries of the expected Fisher information in a 
multivariate elliptical distribution was, perhaps, \cite{Mitchell1989}. Other recent papers 
have adopted the same strategy, namely \cite{Savalli2006} and \cite{Osorioetal2007}. 
 Since writing a matrix by entering element by element is not an efficient way to do it, 
 we present a matrix version of these quantities (considering a more general model) in which, 
besides an aesthetic improvement, one can use it for avoiding that cumbersome task. Moreover, the compactness 
of the expressions might encourage other researches to study more complex models. 
 We also show matrix versions of some expectations of a variable with elliptic distribution that 
can be useful to apply in a multivariate context.

The nonlinear model studied in 
\cite{Russoetal2009} is given by
\begin{equation}\label{model}
\bm{y}_i = \bm{f}(\bm{x}_i, \bm{\alpha}) + \bm{Z}_i\bm{b}_i + \bm{\epsilon}_i, \quad i=1, \ldots, n
\end{equation} and, as defined by the authors, $\bm{f}$ is an $m_i$-dimensional nonlinear function of $\bm{\alpha}$,
$\bm{x}_i$ is a vector of covariates, $\bm{Z}_i$ is a matrix of known constants, 
$\bm{\alpha}$ is a $p \times 1$ vector of unknown parameters and $\bm{b}_i$
is an $r \times 1$ vector of unobserved random regression coefficients, where
$(\bm{y}_i, \bm{b}_i)$ follows an elliptical distribution, such that
\[
\begin{pmatrix}
\bm{y}_i\\
\bm{b}_i
\end{pmatrix} \stackrel{ind}{\sim} \mbox{El}_{m_i+r}
\left[
\begin{pmatrix}
\bm{f}(\bm{x}_i, \bm{\alpha})\\
\bm{0}
\end{pmatrix};
\begin{pmatrix}
\bm{Z}_i\bm{D}\bm{Z}_i^{\top} + \sigma^2\bm{I}_{m_i} & \bm{Z}_i\bm{D}\\
\bm{D}\bm{Z}_i^{\top} & \bm{D}
\end{pmatrix}
\right],
\] where $\bm{I}_{m_i}$ is an ($m_i \times m_i$) identity matrix.

For the purpose of avoiding numerical integrations, \cite{Russoetal2009} consider the marginal model, that is 
$\bm{y}_i \stackrel{ind}{\sim} \mbox{El}_{m_i}(\bm{f}(\bm{x}_i, \bm{\alpha});\bm{\Sigma}_i)$, where 
$\bm{\Sigma}_i=\bm{Z}_i\bm{D}\bm{Z}_i^{\top} + \sigma^2\bm{I}_{m_i}$. The vector of parameters
of interest is defined as $\bm{\theta} = (\bm{\alpha}^{\top}, \bm{\gamma}^{\top})^{\top}$, where
$\bm{\gamma} = (\gamma_0, \gamma_1, \ldots, \gamma_q)^{\top}$ is the vector of parameters involved in 
$\bm{\Sigma}_i$ with, in this case, $\gamma_0 = \sigma^2$. In addition to the authors' suppositions, 
the functional form of $\bm{f}(\bm{x}_i,\bm{\alpha})$ must be known and twice continuously 
differentiable with respect to each element of $\bm{\alpha}$. 
 
In this paper, we consider the following model,
\begin{equation}\label{General}
\bm{y}_i \stackrel{ind}{\sim} \mbox{El}_{m_i}(\bm{f}(\bm{x}_i, \bm{\alpha});\bm{\Sigma}_i(\bm{w}_i,\bm{\gamma})),
\end{equation} where $\bm{w}_i$ and $\bm{x}_i$ may have common components.  The functional form of 
the covariance matrix $\bm{\Sigma}_i(\bm{w}_i,\bm{\gamma})$ is known and twice continuously differentiable with respect to each element of
$\bm{\gamma}$. Since $\bm{\theta}$ must be identifiable in model (\ref{model}), 
we suppose that the model fulfills this requirement. To see that model (\ref{model}) is a special case of
(\ref{General}), take $\bm{w}_i = \bm{Z}_i$ and $\bm{\Sigma}_i(\bm{Z}_i,\bm{\gamma}) = 
\bm{Z}_i\bm{D}\bm{Z}_i^{\top} + \sigma^2\bm{I}_{m_i}$.
As model (\ref{General}) is not considering a specific structure for $\bm{\Sigma}_i$, it can 
represent other multivariate models. That is, model (\ref{model}) can be generalized just by considering
$\bm{R}_i(\bm{z}_i,\bm{\sigma}_2)$ instead of $\sigma^2\bm{I}_{m_i}$, where $\bm{z}_i$ is a vector of extra dispersion covariates. 
Then, in this context, we have that 
$\bm{\Sigma}_i(\bm{w}_i,\bm{\gamma}) = \bm{Z}_i \bm{D}\bm{Z}_i^{\top} + \bm{R}_i(\bm{z}_i,\bm{\sigma})$ and 
$\bm{\gamma} = (\bm{\tau}^{\top},\bm{\sigma}^{\top})^{\top}$, where $\bm{w}_i = (\bm{Z}_i^{\top}, \bm{z}_i^{\top})^{\top}$,
$\bm{\tau}$ is a $q_1 \times 1$ vector of dispersion parameters involved in $\bm{D}$ and $\bm{\sigma}$ is a $q_2 \times 1$
vector of dispersion parameters associated with the model error term. We can go further and assign, for instance, 
a first-order autoregressive covariance matrix to the error terms, that is, $\bm{\Sigma}_i(\bm{w}_i,\bm{\gamma}) = 
\bm{Z}_i \bm{D}\bm{Z}_i^{\top} + \sigma^2\bm{V}(\rho)$, where $\bm{V}_{rs}(\rho) = \rho^{|r-s|}/(1-\rho)$, 
then $\bm{w}_i = \bm{Z}_i$, $q_2 = 2$ and $\bm{\gamma} = (\bm{\tau}, \sigma^2,\rho)^{\top}$. In general, $\bm{\Sigma}_i(\bm{w}_i,\bm{\gamma})$
may be any structured covariance matrix with properties aforementioned. To keep the same notation, consider
 $\bm{\gamma} = (\gamma_0, \ldots, \gamma_q)^{\top}$, i.e., $q_1+q_2 = q+1$, then, the number of 
parameters is still $b=p+q+1$ (here, $b$ is fixed and $b \ll n$).

\cite{Russoetal2009} show that the score functions considering model (\ref{model}) are given by
\begin{equation*}
\bm{U}_{\bm{\alpha}} = \sum_{i=1}^n v_i\bm{J}_i^{\top}\bm{\Sigma}_i^{-1}\bm{r}_i 
\quad \mbox{and} \quad
\bm{U}_{\gamma_j} = -\frac{1}{2}\sum_{i=1}^n 
\left\{\mbox{tr}\big[\bm{\Sigma}_i^{-1}\dot{\bm{\Sigma}}_{i(j)}\big] - v_i \bm{r}_i^{\top}\bm{\Sigma}_i^{-1}\dot{\bm{\Sigma}}_{i(j)}^{-1}\bm{\Sigma}_i^{-1}\bm{r}_i
\right\} \quad \mbox{for} \ j=0,\ldots, q
\end{equation*} where $v_i = -2W_g(u_i)$, $u_i =\bm{r}_i^{\top}\bm{\Sigma}_i^{-1}\bm{r}_i $, 
$\bm{r}_i = \bm{y}_i - \bm{f}(\bm{x}_i, \bm{\alpha})$, 
$\bm{J}_i = \partial \bm{f}(\bm{x}_i, \bm{\alpha})/\partial \bm{\alpha}^{\top}$, 
$\dot{\bm{\Sigma}}_{i(j)} = \partial \bm{\Sigma}_i/\partial\gamma_j$, $W_g(u_i) = d \log g(u_i)/d u_i$
and function $g(\cdot)$ is the density generator function with properties defined in \cite{Russoetal2009}.
 Notice that, the score function $\bm{U}_{\gamma_j}$ has a typographical error. The right form is given by
 \[
 \bm{U}_{\gamma_j} = -\frac{1}{2}\sum_{i=1}^n 
\left\{\mbox{tr}\big[\bm{\Sigma}_i^{-1}\dot{\bm{\Sigma}}_{i(j)}\big] - v_i \bm{r}_i^{\top}\bm{\Sigma}_i^{-1}\dot{\bm{\Sigma}}_{i(j)}\bm{\Sigma}_i^{-1}\bm{r}_i
\right\} \quad \mbox{for} \ j=0,\ldots, q.
 \]
 
 The authors also show that the expected Fisher information considering model (\ref{model}) is given by
\[
\bm{K}_{\bm{\theta\theta}} = 
\begin{pmatrix}
\bm{K}_{\bm{\alpha\alpha}} & \bm{0}\\
\bm{0} & \bm{K}_{\bm{\gamma\gamma}}
\end{pmatrix},
\] where
\[
\bm{K}_{\bm{\alpha\alpha}}  = \sum_{i=1}^n \frac{4d_{gi}}{m_i} \bm{J}_i^{\top}\bm{\Sigma}_i^{-1}\bm{J}_i,
\] and the ($r,s$) element of $\bm{K}_{\bm{\gamma\gamma}}$ is given by
\begin{equation*}\label{Fisher-gamma}
\bm{K}_{\gamma_r\gamma_s}  = \sum_{i=1}^n\left\{ \frac{a_{rsi}}{4}(c_i -1) + c_i\frac{1}{2}
\mbox{tr}\big[\bm{\Sigma}_i^{-1}\dot{\bm{\Sigma}}_{i(r)}\bm{\Sigma}_i^{-1}\dot{\bm{\Sigma}}_{i(s)}\big]\right\}
\end{equation*} with $c_{i} = 4f_{gi}/[m_i(m_i+2)]$ and the quantities $d_{gi}$, $f_{gi}$ and $a_{rsi}$ are well 
defined in \cite{Russoetal2009}. Note that, the above score functions and Fisher information are essentially the same
 of those under model (\ref{General}), but here matrix $\bm{\Sigma}_i$ does not have the specific structure as regarded 
in \cite{Russoetal2009}, it is left in the general form. See also that the score 
function and Fisher information for $\bm{\gamma}$ are written in an element-by-element form.

This paper is organized as follows. Section \ref{Score-Fisher} presents a matrix version for the score function, 
the (observed and expected) Fisher information and shows an iterative re-weighted least squares algorithm to attain 
the maximum-likelihood estimate for $\bm{\theta}$.
Section \ref{Deltas} shows a matrix version for the $\bm{\Delta}$ matrices presented by \cite{Russoetal2009} which are also applicable 
for model (\ref{General}). Additionally, Section \ref{Levarage} presents an expression for the 
generalized leverage in model (\ref{General}). We do not present an application in this paper, since it can be seen
just as a complementary material of \cite{Russoetal2009}.

\subsection{Matrix version for the score function and Fisher information}
\label{Score-Fisher}

The following two matrix results will be intensively used in the computation
of the expressions derived in this paper. Let $\bm{A}$, $\bm{B}$,
 $\bm{C}$ and $\bm{D}$ be $n\times n$ matrices, define also $\bm{A} = (\bm{a}_1, \bm{a}_2, \ldots, \bm{a}_n)$ and 
 $\bm{C} = (\bm{c}_1, \bm{c}_2, \ldots, \bm{c}_n)$, where $\bm{a}_i$ and $\bm{c}_i$ are
 $n\times 1$  vectors, then
 \begin{equation}\label{matrix}
 \mbox{tr}\{\bm{A}^{\top}\bm{C}\bm{D}\bm{B}^{\top}\} = \vvec(\bm{A})^{\top}(\bm{B}\otimes \bm{C})\vvec(\bm{D})
 \quad \mbox{and} \quad  \bm{A}^{\top}\bm{B}\bm{C} = \{\bm{a}_r^{\top}\bm{B}\bm{c}_s\}
 \end{equation}  where vec$(\cdot)$ is the vec operator, which transforms a matrix into a 
vector by stacking the columns of the matrix one underneath the other, ``$\otimes$''
 indicates the Kronecker product. These results and other methods in matrix differential calculus can be studied in
\cite{MagnusNeudecker}.

Define the following quantities,
\[
\bm{F}_i = 
\begin{pmatrix}
\bm{J}_i & \bm{0}\\
\bm{0}   &\bm{V}_i
\end{pmatrix}, \quad 
\bm{H}_i=
\begin{pmatrix}
\bm{\Sigma}_i^{-1} & \bm{0}\\
\bm{0} & \frac{1}{2}\bm{\Sigma}_i^{-1}\otimes\bm{\Sigma}_i^{-1}
\end{pmatrix}, \quad 
\dot{\bm{u}}_i=
\begin{pmatrix}
v_i\bm{r}_i\\
-\vvec(\bm{\Sigma}_i - v_i\bm{r}_i\bm{r}_i^{\top})
\end{pmatrix}
\] and $\bm{V}_i = \big(\vvec(\dot{\bm{\Sigma}}_{i(0)}), \ldots,\vvec(\dot{\bm{\Sigma}}_{i(q)}) \big)$, 
where $\bm{F}_i$ has rank $b$ (i.e., the functions $\bm{f}$ and $\bm{\Sigma}_i$ must be defined to hold such condition). Then, by using (\ref{matrix}) and after a somewhat algebra, we have that the score function 
and the expected Fisher information, considering model (\ref{General}), can be written, respectively, as
\begin{equation}\label{UK}
\bm{U}_{\bm{\theta}} = \sum_{i=1}^n \bm{F}_i^{\top}\bm{H}_i\dot{\bm{u}}_i
\quad \mbox{and} \quad 
\bm{K}_{\bm{\theta\theta}} = \sum_{i=1}^n \bm{F}_i^{\top}\bm{H}_i\bm{O}_i\bm{H}_i\bm{F}_i
\end{equation} where
\[
\bm{O}_i = c_i
\begin{pmatrix}
\frac{4d_{gi}}{m_ic_i}\bm{\Sigma}_i & \bm{0}\\
\bm{0} &   2\bm{\Sigma}_i\otimes\bm{\Sigma}_i
\end{pmatrix} + (c_i -1)
\begin{pmatrix}
\bm{0} & \bm{0}\\
\bm{0} &  \vvec(\bm{\Sigma}_i)\vvec(\bm{\Sigma}_i)^{\top}
\end{pmatrix}.
\]

Fisher information given in (\ref{UK}) can clearly be interpreted as a quadratic form
which can be easily attained through direct matrix operations. 
 Thus, a joint iterative procedure for  attaining the MLE of $\bm{\theta}$ can be formulated as 
the following re-weighted least squares algorithm 
\begin{equation}\label{iterative}
\widehat{\bm{\theta}}^{(m+1)} = 
\left(\sum_{i=1}^n\bm{F}_i^{(m)\top}\widetilde{\bm{H}}_i^{(m)}\bm{F}_i^{(m)}\right)^{-1}
\left(\sum_{i=1}^n \bm{F}_i^{(m)\top}\widetilde{\bm{H}}_i^{(m)}\widetilde{\bm{u}}_i^{(m)}\right), \quad m =1,2, \ldots
\end{equation} where the quantities with the upper script ``$(m)$'' are evaluated at $\widehat{\bm{\theta}}^{(m)}$, 
$\widetilde{\bm{H}}_i = \bm{H}_i\bm{O}_i\bm{H}_i$, 
$\widetilde{\bm{u}}_i = \bm{H}_i^{-1}\bm{O}_i^{-1}\dot{\bm{u}}_i + \bm{F}_i\widehat{\bm{\theta}}$ and $m$ is 
the iteration counter. Under normality we have that $c_i =1$, $\bm{O}_i = \bm{H}_i^{-1}$ and 
$v_i =1$, and it is easy to see that this iterative procedure (under normality) is a special case
 of the one proposed in \cite{Patriota2009}. 



In the sequence, we provide a matrix formulation for the observed Fisher information which requires
harder matrix operations than the one spent in the expected Fisher information. The 
observed Fisher information presented in \cite{Russoetal2009}, that is the same observed Fisher information
considering model~(\ref{General}), is given by $-\ddot{\bm{L}}_{\bm{\theta\theta}} = 
- \sum_{i=1}^n \ddot{\bm{L}}_{\bm{\theta\theta},i}$, with
\[
 \ddot{\bm{L}}_{\bm{\theta\theta},i} =\frac{\partial L_i(\bm{\theta})}{\partial \bm{\theta}\partial \bm{\theta}^{\top}}=
\begin{pmatrix}
\ddot{\bm{L}}_{\bm{\alpha\alpha},i} & \ddot{\bm{L}}_{\bm{\alpha\gamma},i}\\
\ddot{\bm{L}}_{\bm{\gamma\alpha},i} & \ddot{\bm{L}}_{\bm{\gamma\gamma},i}
\end{pmatrix}
\] where
\[
\ddot{\bm{L}}_{\bm{\alpha\alpha},i} = 2\bm{J}_i^{\top} \bm{\Sigma}_i^{-1}\left\{ W_g(u_i)\bm{\Sigma}_i + 2 W_g'(u_i)\bm{r}_i\bm{r}_i^{\top}\right\}\bm{\Sigma}_i^{-1}\bm{J}_i -2W_g(u_i)[ \bm{I}_p \otimes \bm{r}_i^{\top}\bm{\Sigma}_i^{-1}]\bm{D}_i, 
\]
\begin{equation}\label{L-alpha.gamma}
\ddot{\bm{L}}_{\bm{\alpha\gamma},i} = (\ddot{\bm{L}}_{\bm{\alpha}\gamma_0,i}, \ddot{\bm{L}}_{\bm{\alpha}\gamma_1,i}, \ldots, \ddot{\bm{L}}_{\bm{\alpha}\gamma_q,i})
\end{equation} with $\ddot{\bm{L}}_{\bm{\alpha}\gamma_j,i}= 2\bm{J}_i^{\top} \bm{\Sigma}_i^{-1}\left\{ W_g(u_i)\bm{\Sigma}_i +  W_g'(u_i)\bm{r}_i\bm{r}_i^{\top}\right\}\bm{\Sigma}_i^{-1}\dot{\bm{\Sigma}}_{i(j)}\bm{\Sigma}_i^{-1}\bm{r}_i$ and the element 
($j,k$) of $\ddot{\bm{L}}_{\bm{\gamma\gamma},i}$ has the form 
\begin{align}\label{L-gamma.gamma}
\frac{1}{2}\mbox{tr}\left\{\bm{\Sigma}_i^{-1}\bigg[\dot{\bm{\Sigma}}_{i(j)}\bm{\Sigma}_i^{-1}\dot{\bm{\Sigma}}_{i(k)} -\dot{\bm{\Sigma}}_{i(jk)}\bigg]  \right\} &+ \bm{r}_i^{\top}\bm{\Sigma}_i^{-1}\left\{ W_g'(u_i)\dot{\bm{\Sigma}}_{i(j)}\bm{\Sigma}_i^{-1}\bm{r}_i\bm{r}_i^{\top}\bm{\Sigma}_i^{-1}\dot{\bm{\Sigma}}_{i(k)} -
 W_g(u_i)\dot{\bm{\Sigma}}_{i(jk)}\right.\\\nonumber
 &\left. 
 + W_g(u_i)\dot{\bm{\Sigma}}_{i(j)}\bm{\Sigma}_i^{-1}\dot{\bm{\Sigma}}_{i(k)} + W_g(u_i)\dot{\bm{\Sigma}}_{i(k)}\bm{\Sigma}_i^{-1}\dot{\bm{\Sigma}}_{i(j)}\right\}
 \bm{\Sigma}_i^{-1}\bm{r}_i
 \end{align}%
with
\[
\dot{\bm{\Sigma}}_{i(jk)} = \frac{\partial^2 \bm{\Sigma}_i}{\partial \gamma_j \partial \gamma_k}, \qquad
\bm{D}_i=
\begin{pmatrix}
\bm{a}_{i(11)} &  \ldots & \bm{a}_{i(1p)}\\
\vdots         &  \ddots & \vdots\\
\bm{a}_{i(p1)} &  \ldots & \bm{a}_{i(pp)}\\
\end{pmatrix}\qquad \mbox{and} \qquad \bm{a}_{i(rs)} = \frac{\partial^2 \bm{f}}{\partial \alpha_r\partial \alpha_s}.\] 

Note that, quantities (\ref{L-alpha.gamma}) and (\ref{L-gamma.gamma}) are not written 
in a matrix form, in the following we present a compact matrix version of $\ddot{\bm{L}}_{\bm{\theta\theta}}$.
\begin{equation}\label{ObsF}
\ddot{\bm{L}}_{\bm{\theta\theta}} = \sum_{i=1}^n \left\{ \bm{F}_i^{\top}\bm{H}_i\ddot{\bm{O}}_i\bm{H}_i\bm{F}_i +
\left[\dot{\bm{u}}_i^{\top}\bm{H}_i\right]\left[\frac{\partial \bm{F}_i}{\partial \bm{\theta}}\right]\right\}
\end{equation} where 
\begin{align*}
\ddot{\bm{O}}_i = 2W_g(u_i)&
\begin{pmatrix}
\bm{\Sigma}_i                & 2\bm{\Sigma}_{i} \otimes \bm{r}_{i}^{\top}\\
2\bm{\Sigma}_{i}\otimes \bm{r}_i & 2(\bm{\Sigma}_i\otimes(\bm{r}_i\bm{r}_i^{\top}) +(\bm{r}_i\bm{r}_i^{\top})\otimes \bm{\Sigma}_i )
\end{pmatrix} +2
\begin{pmatrix}
\bm{0}& \bm{0}\\
\bm{0}& \bm{\Sigma}_i\otimes \bm{\Sigma}_i,
\end{pmatrix}\\ 
& + 4W_g'(u_i)
\begin{pmatrix}
\bm{r}_i\bm{r}_i^{\top} & (\bm{r}_i\bm{r}_i^{\top})\otimes \bm{r}_i^{\top}\\
(\bm{r}_i\bm{r}_i^{\top})\otimes \bm{r}_i  & \vvec(\bm{r}_i\bm{r}_i^{\top})\vvec(\bm{r}_i\bm{r}_i^{\top})^{\top}
\end{pmatrix},
\end{align*}
$\frac{\partial \bm{F}_i}{\partial \bm{\theta}}$ is an $m_i(m_i+1)\times b \times b$ 
array, 
$\left[\dot{\bm{u}}_i^{\top}\bm{H}_i\right]\left[\frac{\partial \bm{F}_i}{\partial \bm{\theta}}\right]$ is 
the bracket product of $\dot{\bm{u}}_i^{\top}\bm{H}_i$ and $\frac{\partial \bm{F}_i}{\partial \bm{\theta}}$ 
 \cite[for further details see][on pg. 188]{Wei98}.

In what follows, we present some matricial results on elliptical variables. 
Here, $\bm{r}_i \stackrel{ind}{\sim} \mbox{El}_{m_i}(\bm{0}, \bm{\Sigma}_i)$, then
adapting the results of \cite{Mitchell1989} for a matrix version, we have that

\begin{itemize}

\item[a)] $E(\bm{r}_iv_i) =\bm{0}$, 

\item[b)] $E(\bm{r}_i\bm{r}_i^{\top}v_i) = \bm{\Sigma}_i$, 

\item[c)]$E(\bm{r}_i\bm{r}_i^{\top}v_i^2) = 4d_{gi}/m_i \bm{\Sigma}_i$

\item[d)] $E(\vvec(\bm{r}_i\bm{r}_i^{\top})\bm{r}_i^{\top}v_i^2) = \bm{0}$

\item[e)] $E(\vvec(\bm{r}_i\bm{r}_i^{\top})\vvec(\bm{r}_i\bm{r}_i^{\top})^{\top} v_i^2) = 
c_i\bigg(\vvec(\bm{\Sigma}_i)\vvec(\bm{\Sigma}_i)^{\top} + \bm{\Sigma}_i\otimes \bm{\Sigma}_i + \bm{P}_i(\bm{\Sigma}_i\otimes \bm{\Sigma}_i)\bigg),$ where $\bm{P}_i$ is a commutation maltrix such that $\mbox{vec}(\bm{A}) = \bm{P}_i \mbox{vec}(\bm{A}^\top)$ for any matrix $\bm{A}$ with appropriated dimensions.

\end{itemize}
 
Therefore, as we are considering a function $g(\cdot)$ with regular properties (differentiation and 
integration are interchangeable), we have that $E(\dot{\bm{u}}_i) = \bm{0}$  and $E(-\ddot{\bm{L}}_{\bm{\theta\theta}}) = {\bm{K}}_{\bm{\theta\theta}}$.

\subsection{Matrix version for $\bm{\Delta}$}\label{Deltas}

The diagnostic technique developed in \citet{Cook1986} is a well-spread tool
to check the model assumptions and conduct diagnostic studies.
The author proposes to look at the likelihood displacement $LD(\bm{\omega})
=2\{L(\widehat{\bm{\theta}}) - L(\widehat{\bm{\theta}}_{\bm{\omega}})\}$
to find possible influential observations in the MLEs, where $L(\bm{\theta})=\sum_i L_i(\bm{\theta})$ is the 
log-likelihood function and $\bm{\omega}$ is a $s\times1$ vector of perturbation
restricted in an open set $\bm{\Omega} \subset \mathbb{R}^{s}$. It is also defined a
vector of no perturbation as $\bm{\omega}_0 \in \bm{\Omega}$ in which $LD(\bm{\omega}_0) = 0$, i.e.,
$ L(\bm{\theta}_{\bm{\omega}_0}) = L(\bm{\theta})$. In his seminal paper,
Cook shows that the normal curvature at the unit direction $\bm{\ell}$ has the following
form $C_{\bm{\ell}}(\bm{\theta}) = 2 |\bm{\ell}^{\top} \bm{\Delta}^{\top}(\ddot{\bm{L}}_{\bm{\theta\theta}})^{-1} \bm{\Delta}\bm{\ell}|$
where $\bm{\Delta} = \partial^2L(\bm{\theta}|\bm{\omega})
/ \partial\bm{\theta} \partial\bm{\omega}^{\top}$, both  $\bm{\Delta}$ and 
$\Ddot{\bm{L}}_{\bm{\theta}\bm{\theta}}$ are evaluated at 
$\bm{\theta} = \widehat{\bm{\theta}}$ and $\bm{\omega} = \bm{\omega}_{0}$. 
Thus, $C_{\bm{d}_{\max}}$ is twice the largest eigenvalue of $\bm{B} =
-\bm{\Delta}^{\top}\ddot{\bm{L}}_{\bm{\theta}\bm{\theta}}^{-1}\bm{\Delta}$
and $\bm{d}_{\max}$ is the corresponding eigenvector. The index plot of $\bm{d}_{\max}$ 
may reveal how to perturb the model (or data) to obtain large changes in the estimate of
$\bm{\theta}$. For a more detailed information, we refer the reader to the work of \cite{Russoetal2009} and the references
therein. 

Note that, by using the defined quantities, we can write the $b\times n$ matrix $\bm{\Delta}$ in the case 
weight perturbation (i.e., $L_i(\bm{\theta}_{\bm{\omega}_0}) = \omega_iL_i(\bm{\theta})$) and the scale perturbation
(i.e., the perturbed log-likelihood function $L_i(\bm{\theta}_{\bm{\omega}})$ is built replacing $\bm{\Sigma}_i$ with $\omega_i^{-1}\bm{\Sigma}_{i}$ in $L_i(\bm{\theta})$), respectively, by
\begin{equation}\label{Delta1}
\bm{\Delta} = \bigg(\widehat{\bm{F}}_1^{\top}\widehat{\bm{H}}_1\widehat{\dot{\bm{u}}}_1, \ldots, \widehat{\bm{F}}_n^{\top}\widehat{\bm{H}}_n\widehat{\dot{\bm{u}}}_n\bigg)
\quad \mbox{and} \quad \bm{\Delta} = \bigg(\widehat{\bm{F}}_1^{\top}\widehat{\bm{H}}_1\widehat{\dot{\bm{v}}}_1, \ldots, \widehat{\bm{F}}_n^{\top}\widehat{\bm{H}}_n\widehat{\dot{\bm{v}}}_n\bigg),
\end{equation} where the quantities with ``\ $\widehat{}$ '' are evaluated at $\widehat{\bm{\theta}}$ and
\[
\dot{\bm{v}}_i=-2(W_g(u_i) + u_i W_g'(u_i))
\begin{pmatrix}
\bm{r}_i\\
\vvec(\bm{r}_i\bm{r}_i^{\top})\\
\end{pmatrix}.
\] In \cite{Russoetal2009}, the $\bm{\Delta}$ matrix under a case weight perturbation is presented 
with the same typo of the score function. Finally, the $b\times N$ matrix 
 $\bm{\Delta}$ under the response perturbation (i.e., the 
perturbed log-likelihood function $L_i(\bm{\theta}_{\bm{\omega}})$ is built replacing $\bm{y}_i$ with $\bm{y}_i + \bm{\omega}_i$ in $L_i(\bm{\theta})$) becomes
\begin{equation}\label{Delta2}
\bm{\Delta} = \bigg(\widehat{\bm{F}}_1^{\top}\widehat{\bm{H}}_1\widehat{\bm{G}}_1, \ldots, \widehat{\bm{F}}_n^{\top}\widehat{\bm{H}}_n\widehat{\bm{G}}_n\bigg),
\end{equation} where $N = \sum_{i=1}^n m_i$ and
\[
\bm{G}_i =-2
\begin{pmatrix}
W_g(u_i)\bm{I}_{m_i} + 2W'_g(u_i)\bm{r}_i\bm{r}_i^{\top}\bm{\Sigma}_i^{-1} \\
2 \bm{r}_i\otimes \big(W_g(u_i)\bm{I}_{m_i} + W'_g(u_i)\bm{r}_i\bm{r}_i^{\top}\bm{\Sigma}_i^{-1}\big)
\end{pmatrix}.
\]

Note that, formulas (\ref{Delta1}) and (\ref{Delta2}) are easily handled through any 
statistical software. 

\subsection{Generalized leverage}\label{Levarage}

In this section, we compute the generalized leverage proposed by \cite{WeiHuFung1998}.
 Let $\bm{y} = \text{vec}(\bm{y}_1, \ldots, \bm{y}_n)$ and
$\bm{\mu}(\bm{\alpha}) = \text{vec}(\bm{f}(\bm{\alpha}, \bm{x}_1), \ldots,\bm{f}(\bm{\alpha}, \bm{x}_n))$.
 The authors have shown that the generalized leverage is obtained by
evaluating the $N\times N$ matrix 
\[
\bm{GL}(\bm{\theta}) =
\bm{D}_{\bm{\theta}}(-\ddot{\bm{L}}_{\bm{\theta\theta}})^{-1}
\ddot{\bm{L}}_{\bm{\theta}\bm{Y}}, 
\]
at $\bm{\theta} = \widehat{\bm{\theta}}$, where
$\bm{D}_{\bm{\theta}} = \partial\bm{\mu}(\bm{\alpha})/\partial\bm{\theta}^{\top}$
and $\ddot{\bm{L}}_{\bm{\theta}\bm{Y}} = \partial^2\ell(\bm{\theta})
/\partial\bm{\theta}\partial\bm{Y}^{\top}$.
The main idea behind the concept of leverage is that of 
evaluating the influence of $\bm{Y}_{i}$ on its own predicted
value. As noted by the authors, the generalized leverage is invariant under
reparameterizations and observations with large $GL_{ii}$ are leverage
points. 

Under the model defined in~(\ref{General}), we have that
\[
\bm{D}_{\bm{\theta}} = 
\begin{pmatrix}
\bm{J}_1& \bm{0}\\
\bm{J}_{2}& \bm{0}\\
\vdots & \vdots \\
\bm{J}_{n}& \bm{0}
\end{pmatrix} \qquad \mbox{and} \qquad
\ddot{\bm{L}}_{\bm{\theta}\bm{Y}} = \Bigl(\bm{F}_1^{\top}\bm{H}_1\bm{G}_{1},\ldots, 
\bm{F}_n^{\top}\bm{H}_n\bm{G}_{n}\Bigr)
\]

Index plots of $GL_{ii}$ may reveal those observations with high influence on their 
own predicted values.

It is worth emphasizing that other models are special cases of the formulas derived in this paper. 
One just has to define $\bm{f}(\bm{x}_i,\bm{\alpha})$ and $\bm{\Sigma}_i(\bm{w}_i,\bm{\gamma})$ 
 and find their derivatives. That is, the score vector and the (expected and observed) Fisher 
 information as well as the curvatures and the generalized leverage (when available) of several 
 works are special cases of the proposed matrix formulation \citep[to mention just a few of 
 them, see for instance,][]{Paulaetal2003, Savalli2006, Osorioetal2007, Paulaetal2009, Russoetal2009}.

\section{Conclusion}

In this short communication, we presented  a matrix formulation of the score function, the 
(expected and observed) Fisher information, the generalized leverage and the $\bm{\Delta}$ matrices under case 
weight, scale and response perturbations for a very general elliptical model which 
includes the nonlinear mixed-effects elliptical model proposed in \cite{Russoetal2009}. 
 The general expressions derived in this paper can be applied in many other models 
and have advantages for numerical purposes because they require only simple operations on matrices 
and vectors.
 
\section*{Acknowledgments}

I gratefully acknowledge grants from FAPESP.

\end{document}